%% file: main.tex
\documentclass[acmsmall,screen]{acmart}
\usepackage{xspace}

\usepackage{amsmath,amsfonts}
\usepackage{graphicx}
\usepackage{tabularx}
\usepackage{booktabs} 
\usepackage{stfloats}
\usepackage{caption}  
\usepackage{textcomp}
\usepackage{xcolor}
\usepackage{makecell}
\usepackage{enumitem}
\usepackage{balance}
\usepackage{url}
\usepackage[splitrule]{footmisc}

\usepackage[most]{tcolorbox}
\tcbuselibrary{listings, breakable}
\usepackage{xcolor}
\usepackage{xurl}

\lstdefinelanguage{CodeQL}{
  morekeywords={from,select,where,import,class,predicate,module},
  sensitive=true,
  morecomment=[l]{//},
  morestring=[b]",
}

\definecolor{codegreen}{rgb}{0,0.6,0}
\definecolor{codegray}{rgb}{0.5,0.5,0.5}
\definecolor{codepurple}{rgb}{0.58,0,0.82}

\begin{document}

\title{Towards Scalable and Cost-Efficient Vulnerability Detection: \\A Study on Automatic Query Generation}

\author{Ivana Clairine Irsan}
\email{ivanairsan@smu.edu.sg}
\affiliation{
  \institution{Singapore Management University}
  \country{Singapore}
}

\author{Ratnadira Widyasari}
\email{ratnadiraw@smu.edu.sg}
\affiliation{
  \institution{Singapore Management University}
  \country{Singapore}
}

\author{Huihui Huang}
\email{hhhuang@smu.edu.sg}
\affiliation{
  \institution{Singapore Management University}
  \country{Singapore}
}

\author{Ting Zhang}
\email{Ting.Zhang@monash.edu}
\affiliation{
  \institution{Monash University}
  \country{Australia}
}

\author{Yue Liu}
\email{liuyue@smu.edu.sg}
\affiliation{
  \institution{Singapore Management University}
  \country{Singapore}
}

\author{Ouh Eng Lieh}
\email{elouh@smu.edu.sg}
\affiliation{
  \institution{Singapore Management University}
  \country{Singapore}
}

\author{Shar Lwin Khin}
\email{lkshar@smu.edu.sg}
\affiliation{
  \institution{Singapore Management University}
  \country{Singapore}
}

\author{Kang Hong Jin}
\email{hongjin.kang@sydney.edu.au}
\affiliation{
  \institution{University of Sydney}
  \country{Australia}
}

\author{David Lo}
\email{davidlo@smu.edu.sg}
\affiliation{
  \institution{Singapore Management University}
  \country{Singapore}
}



\begin{abstract}
Static analysis remains a cornerstone of software security, yet the effectiveness of tools such as CodeQL is often limited by the substantial manual effort required to develop high-coverage query suites. 
While large language models (LLMs) have emerged as a potential solution for automated code reasoning, their practical utility in generating structured, executable security queries remains underexplored. 
In this paper, we conduct an empirical study to evaluate the ability of LLMs to synthesize CodeQL queries using vulnerability data from the National Vulnerability Database.
Through this investigation, we explore the potential of using LLMs as an automatic CodeQL query generator.
Subsequently, we systematically evaluate the performance of various LLM architectures across a diverse set of real-world vulnerabilities, measuring their ability to improve detection coverage and precision.
Our findings reveal that LLM-generated queries significantly enhance the baseline CodeQL queries, yielding 82\% improvement in average F1-score.
Furthermore, we provide a detailed cost-benefit analysis showing that while direct LLM-based scanning of entire repositories is often computationally and financially prohibitive, leveraging LLMs to synthesize CodeQL queries offers a scalable and cost-effective alternative for large-scale vulnerability detection.
Our results suggest that LLMs can effectively bridge the gap between unstructured vulnerability reports and formal static analysis specifications, offering a scalable path toward comprehensive automated vulnerability detection.
\end{abstract}

\maketitle

\newcommand{\codeql}{{CodeQL}\xspace}

\newboolean{showcomments}
\setboolean{showcomments}{false}
\ifthenelse{\boolean{showcomments}}
 { \newcommand{\mynote}[2]{
      \fbox{\bfseries\sffamily\scriptsize#1}
        {\small$\blacktriangleright$\textsf{\emph{#2}}$\blacktriangleleft$}}}
        { \newcommand{\mynote}[2]{}}

\newcommand{\todoc}[2]{{\textcolor{#1} {\textbf{#2}}}}
\newcommand{\todo}[1]{{\todoc{red}{\textbf{#1}}}}
\newcommand{\iv}[1]{\mynote{Ivana}{\todoc{blue}{#1}}}
\newcommand{\rw}[1]{\mynote{Ratna}{\todoc{violet}{#1}}}
\newcommand{\hj}[1]{\mynote{Hong Jin}{\todoc{magenta}{#1}}}
\newcommand{\zt}[1]{\mynote{Ting}{\todoc{olive}{#1}}}
\newcommand{\ly}[1]{\mynote{Yue}{\todoc{brown}{#1}}}
\newcommand{\rev}[1]{{\leavevmode\color{teal}{#1}}}

\newtcolorbox{querybox}[1][]{
  enhanced,
  breakable,
  colback=gray!5,
  colframe=black!50,
  fonttitle=\bfseries,
  title=#1,
  left=6pt, right=6pt, top=6pt, bottom=6pt,
  boxrule=0.6pt,
  arc=2pt,
  sharp corners=south,
  listing only,
  listing options={style=tcblatex,breaklines=true},
  width=\linewidth
}

\newtcolorbox{rqbox}[1]{
    colback=gray!5,     
    colframe=gray!80,    
    fonttitle=\bfseries,
    title=#1,            
    arc=1mm,             
    boxrule=0.5pt,       
    left=2mm, right=2mm, top=1mm, bottom=1mm, 
    enhanced,
    drop shadow={black!20!white} 
}



\renewcommand{\thetable}{\arabic{table}}

\input{1_introduction.tex}
\input{2_background}

\input{3_framework}

\input{4_evaluation}

\input{5_result}
\input{6_lesson_learned}

\input{7_related_works}

\input{8_future_works}
\input{9_data_availability}

\balance
\bibliographystyle{ACM-Reference-Format}
\bibliography{references}


\end{document}

%% file: 1_introduction.tex
\section{Introduction}

\hj{are we using the right format? we should be using ACM format with sigconf,review,anonymous }


Software vulnerabilities continue to pose severe security risks to modern software systems~\cite{lin2024vulnerabilities}.
High-impact incidents such as Log4Shell~\cite{apacheLog4shell2021} and Spring4Shell~\cite{spring4shell2022} disrupted thousands of applications in Java ecosystems, causing widespread financial damage and demonstrating how a single vulnerability can escalate into a global security crisis. 
Such incidents underscore the critical need for proactive vulnerability detection. 
Rather than reacting after a compromise has occurred, security efforts must learn from historical vulnerabilities to anticipate and prevent their recurrence.

To this end, static application security testing (SAST) tools are widely adopted in industrial software development due to their ability to analyze source code without execution.
This enables early detection in the software development lifecycle (SDLC), significantly reducing remediation costs compared to dynamic testing approaches. 
However, the effectiveness of SAST tools fundamentally depends on manually written vulnerability rules.
These rules must accurately encode complex vulnerability patterns,
yet recent empirical evaluations~\cite{li2023comparison} reveal that SAST tools detect only 12.7\% of real-world vulnerabilities, even when multiple tools are combined. 
Consequently, over 70\% of vulnerabilities remain undetected due to the limited coverage and specificity of existing rule sets.

Among SAST tools, CodeQL~\cite{li2023comparison, szabo2023incrementalizing, shen2025finding} has gained widespread adoption in both academia and industry due to its semantic code representation, which is based on data flow and control flow modeling, and its queryable code database paradigm. CodeQL enables security practitioners to write declarative queries to detect vulnerability patterns. However, writing CodeQL rules is notoriously difficult: it requires expertise in program analysis, familiarity with CodeQL libraries, and a deep understanding of vulnerability semantics. 
This creates a rule-authoring bottleneck, resulting in insufficient detection coverage. 
Meanwhile, large language models (LLMs) have shown strong capabilities in structured code generation, including generating source code, program repair patches, and SQL queries. This raises a compelling question: \textit{Can LLMs help generate effective CodeQL vulnerability queries to expand detection coverage and reduce reliance on manual security expertise?}

In this paper, we explore the potential of bridging the gap between LLMs and static analysis through an empirical study of an automated framework that generates CodeQL queries from NVD data.

Overall, our framework extracts patterns from VFCs and natural language descriptions to guide an LLM in synthesizing executable CodeQL queries via structured prompt engineering and iterative refinement. 
Our evaluation on Java vulnerabilities, which is based on the MITRE ``Top 25 Most Dangerous Software Weaknesses"~\cite{mitre2025cwetop25} demonstrates that LLM-generated queries significantly expand detection coverage.
Specifically, our empirical results demonstrate that LLM-generated queries increased true positive detections by 263\% compared to default CodeQL queries, notably without a significant increase in the False Detection Rate (FDR).

This paper makes the following contributions:
\begin{itemize}[leftmargin=*]
    \item \textbf{Comprehensive Evaluation of LLMs}: We perform an extensive benchmarking of multiple state-of-the-art (SOTA) LLMs, spanning both proprietary and open-source architectures, to evaluate their efficacy in automated CodeQL query synthesis.
    \item \textbf{High-Impact Security Analysis}: We assess the capability of LLMs to generate queries targeting the most dangerous CWE IDs, addressing the practical necessity of securing real-world software against high-risk vulnerability classes.
    \item \textbf{Practicality and Cost-Efficiency Assessment}: We provide a rigorous analysis of the monetary and computational overhead associated with LLM-driven detection. Furthermore, we benchmark our approach against established SAST tools and state-of-the-art function-level deep learning models to evaluate its viability for industrial deployment.
    \item \textbf{Collaborative Vulnerability Detection}: We show that the collaboration between LLM-driven query synthesis and static analysis provides a scalable pathway for future research in autonomous and practical vulnerability detection.
\end{itemize}

%% file: 2_background.tex
\section{Background}\label{sec:background}

\subsection{Static Analyzers}

Static analyzers are tools designed to examine source code, bytecode, or binaries for potential vulnerabilities, bugs, or code quality issues without executing the program. 
They operate by modeling program structures (e.g., control flow and data flow) or by matching predefined patterns, enabling early detection of defects in the SDLC. This early detection can reduce the cost of fixing issues compared to post-deployment remediation.
Among static analyzers, semantic analysis tools that enable inter-procedural vulnerability detection are particularly beneficial for security vulnerability detection.
They can trace complex data flows (e.g., tainted user input propagation) that syntactic tools, which rely on simple pattern matching, often miss.

CodeQL~\cite{li2023comparison, szabo2023incrementalizing, shen2025finding}, a semantic static analysis engine developed by GitHub, exemplifies this capability by treating source code as a relational database. It parses code into an abstract syntax tree (AST), then constructs a queryable database of code elements (e.g., variables, functions, and control flows). Users can then write declarative queries in CodeQL’s logic-based language (similar to SQL) to identify vulnerability patterns—for instance, tracing how untrusted input (sources) reaches sensitive operations (sinks) in data flow analyses. This flexibility makes CodeQL widely adopted in industry and academia for detecting complex vulnerabilities across large codebases, including Java programs with intricate class hierarchies and inter-procedural dependencies. 
We selected CodeQL over alternative SAST tools such as Semgrep due to its native support for whole-program inter-procedural analysis; notably, Semgrep's open-source Community Edition is limited to intra-procedural, single-function analysis by design, and cross-function taint tracking is only available in its proprietary paid tier~\cite{semgrep_docs_interfile}.

However, CodeQL’s power is constrained by the quality of its queries. 
Writing effective CodeQL queries requires two forms of expertise: (1) deep knowledge of the target language’s vulnerability semantics (e.g., how cross-site scripting (XSS) manifests in Java’s server-side code) and (2) proficiency in CodeQL’s query syntax (e.g., defining data flow configurations and filtering false positives). 
These two requirements create a significant barrier of adoption for most developers, who may understand Java security but lack the skills to translate that knowledge into precise CodeQL queries. 
As a result, many organizations only use the prebuilt queries from the official CodeQL repositories. 
However, this leads to critical gaps in vulnerability coverage, as evidenced by recent empirical evaluations of Java static analyzers~\cite{li2023comparison}.



\subsection{Motivation}




The official CodeQL repository provides a set of prebuilt queries for Java vulnerability detection, but these queries fail to cover many common and high-impact vulnerability patterns. 
This limitation is particularly concerning given the prevalence of Java in enterprise systems and the severity of its associated security risks. For example, Cross-Site Scripting (CWE-79), which is ranked among the top five most dangerous software weaknesses in 2025. 
Despite this, CodeQL's standard Java security suite provides only a single prebuilt query for non-Android applications, leaving a significant portion of web-based enterprise vulnerabilities unaddressed.
This query adopts a taint analysis that checks whether user-controlled input (sources, such as \texttt{HttpServletRequest.getParameter()}) flows into sensitive output operations (sinks, such as \texttt{JspWriter.print()}) without proper sanitization. 
While this rule captures basic XSS patterns, it overlooks alternative attack vectors frequently encountered in real-world Java applications.
Specifically, it fails to account for novel sources and sinks documented in recent vulnerability reports, which remain undetected by static, prebuilt query suites.

These limitations indicate a pressing need to enhance CodeQL’s detection coverage by generating queries that address real-world vulnerability patterns beyond simple source-to-sink flows. However, writing such queries manually is challenging and requires deep domain expertise in both Java security and static analysis. In this context, LLMs present a promising opportunity. LLMs have demonstrated strong capabilities in generating structured code and query logic from natural language descriptions, suggesting their potential to support or automate CodeQL query authoring.

On the other hand, directly utilizing LLMs to scan entire repositories for vulnerable code presents significant practical challenges. 
For non-open-source projects, uploading entire codebases to a third-party provider poses a critical security risk by potentially exposing sensitive intellectual property. 
Even for open-source projects where confidentiality is less of a concern, the financial and temporal costs can be prohibitive at the repository level.

Our preliminary experiments on utilizing only the LLM agent as a vulnerability detector highlighted these operational inefficiencies during a full-repository scan of the \texttt{aerospike-client-java} repository\footnote{https://github.com/aerospike/aerospike-client-java}, which comprises 395 files and 3,719 functions. 
We selected this specific project for our cost-analysis baseline by randomly sampling from the subset of evaluations where our LLM-generated queries achieved a high Average F1-Score (more than 0.80). 
Consequently, we utilize this repository as a preliminary benchmark to evaluate the operational costs associated with scanning a single, real-world project.

Using the Moonshot Kimi K2.5 model, which we selected for its optimal performance to cost ratio, the scan required over 44 hours to complete and cost approximately 17 USD for a single iteration.
To identify the most suitable model, a researcher might need to benchmark several LLMs, which could easily exceed ten times the cost of a single Kimi K2.5 iteration. 
For instance, conducting a full repository scan using Sonnet 4.5 would be approximately six times more expensive. When multiplied across several candidate models, these cumulative financial and temporal requirements make direct repository-wide LLM scanning impractical for standard development cycles.

Furthermore, manual sampling revealed multiple false positive alerts, with many non-vulnerable files incorrectly flagged. For larger enterprise-scale repositories, such monetary and time requirements would likely dissuade developers from integrating direct LLM scanning into their workflows.



%% file: 3_framework.tex
\section{Methodology}\label{sec:methodology}

\begin{figure*}[htb!]
    \centering
    \includegraphics[width=1\textwidth]{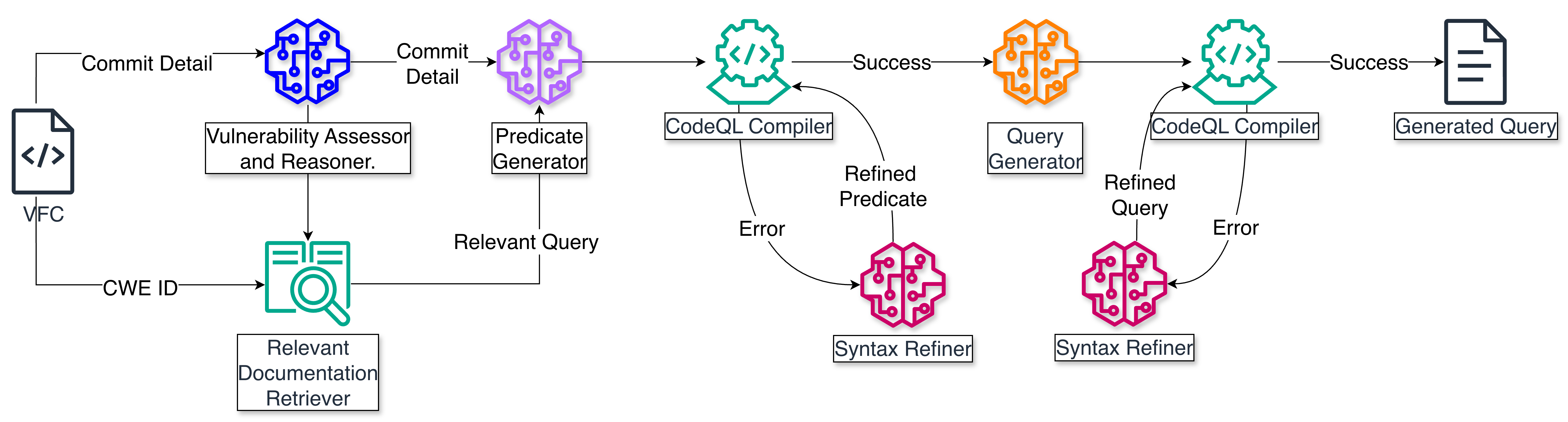}
    \caption{Overview of our query generation pipeline. 
    }
    \label{fig:overview}
\end{figure*}

\subsection{Research Questions}
To better understand the capabilities, limitations, and practical implications of LLM-generated queries, we address the following key research questions (RQs): 

\vspace{4px}
\noindent \textbf{RQ1: How well do LLMs perform in generating compilable \codeql queries?}
CodeQL queries are significantly more complex and have fewer publicly available training examples compared to common languages such as SQL,  meaning LLM exposure to CodeQL may be inherently limited. 
It is therefore essential to evaluate which models are capable of producing syntactically correct, compilable queries before any downstream use.
To address this RQ, we conducted experiments across a diverse set of both open-source and commercial LLMs to measure their performance in generating compilable queries. In this experiment, we include DeepSeek R1~\cite{guo2025deepseek}, Grok Code Fast 1~\cite{xai2025grokcodefast}, Gemini 3 Flash Preview~\cite{google2025gemini3flash}, Kimi ~\cite{team2026kimi}, Kimi K2 Thinking~\cite{team2025kimi}, Llama 3.3~\cite{grattafiori2024llama}, Minimax M2.1~\cite{minimax2025minimaxm1scalingtesttimecompute}, Qwen 3 Coder~\cite{yang2025qwen3technicalreport}, GPT 5.2 Codex~\cite{openai2025gpt52codex}, and Claude Sonnet 4.5~\cite{anthropic2025sonnet45}.

\vspace{4px}
\noindent \textbf{RQ2: How well do the LLM-generated queries perform in detecting real-world vulnerabilities?}
Compilability is a necessary but not sufficient condition for utility; a syntactically valid query may still fail to detect actual vulnerabilities in practice.
To address this RQ, we evaluated the detection performance of LLM-generated queries against CodeQL and PDBERT~\cite{liu2024pre} baselines at file-level granularity on a set of real-world, high-impact software vulnerabilities.

\vspace{4px}
\noindent \textbf{RQ3: Under what conditions do LLM-generated queries perform better or worse?}
Identifying the scenarios in which LLM-generated queries succeed or fail is key to understanding their practical utility and guiding developers on where automation offers the greatest benefit.
To address this RQ, we used CVEs with short fixes as a proxy for intra-procedural vulnerabilities~\cite{11408110} and compared LLM-generated query performance on these localized flaws against inter-procedural vulnerabilities.

\vspace{4px}
\noindent \textbf{RQ4: Which LLM setup provides the best trade-off between effectiveness and cost?}
High operational cost is a practical barrier to adopting LLMs for automated query generation, making it important to identify configurations that balance detection effectiveness with efficiency.
To address this RQ, we evaluated each model's detection performance relative to its operational cost, deriving actionable guidelines for developers building automated CodeQL query generators from existing vulnerability reports.


\subsection{Pipeline Overview}
As motivated in Section~\ref{sec:background}, directly scanning repositories with LLMs is cost-prohibitive, while existing CodeQL queries suffer from limited vulnerability coverage. Our pipeline bridges this gap by automatically generating 
vulnerability-specific CodeQL queries from VFCs (Figure~\ref{fig:overview}), decomposing the task into three phases: \textit{semantic analysis}, \textit{guided synthesis}, and \textit{iterative refinement}.
In the first phase, the \textit{Vulnerability Assessor and Reasoner} evaluates the VFC to determine its suitability for generalization and extracts its core semantic characteristics. 
Subsequently, the \textit{Relevant Documentation Retriever} identifies pertinent CodeQL APIs via a Retrieval-Augmented Generation (RAG) component to provide a grounded context for synthesis. 
In the final phase, our specialized agents, i.e., the \textit{Predicate Generator} and \textit{Query Generator}, incrementally construct and compose reusable logical predicates into a complete query. 
To ensure practical utility, each query is validated by a \textit{Compilability Checker}, with any identified syntactic errors being resolved through iterative correction by the \textit{Syntax Refiner}.
We treat each file within a VFC as a distinct data point. Specifically, we use each modified segment to generate a corresponding CodeQL query, which we define as \textit{CVE-code segment pairs}. Consequently, a single CVE may result in multiple queries if it involves extensive code changes spanning different files.


The interactions within the framework are organized among the following actors:

\begin{enumerate}[leftmargin=*]

\item \textbf{Vulnerability Assessor and Reasoner.}
    This LLM agent is responsible for evaluating whether a given VFC is relevant and generalizable into a detection pattern. 
    It reasons over the commit changes and summarizes the core characteristics of the fixed vulnerability. 
    If a VFC is determined to be irrelevant or lacks a generalizable pattern for mining, it is discarded, and the synthesis pipeline is terminated for that specific instance.

\item \textbf{Relevant Documentation Retriever.}
    The primary objective of this agent is to identify and retrieve the specific CodeQL APIs required for synthesizing predicates and queries. Given the vast library of available CodeQL functions, the agent enhances generation efficiency by constraining the search space to only the most relevant candidates. Specifically, the agent utilizes a CWE-ID as a query for an LLM-based search engine, which is tasked with extracting the standard APIs and predicates typically employed to model and detect that specific vulnerability class.

\item \textbf{CodeQL Predicate Generator.}
    This agent identifies vulnerability patterns from pairs of VFCs and their vulnerable versions. 
    By leveraging retrieved documentation, it translates these patterns into CodeQL predicates (i.e., used to describe logical relations in CodeQL that capture aspects of vulnerability behavior). 
    By generating predicates rather than full queries, the agent can focus on converting unstructured logic into formal code without the overhead of query structural requirements.

\item \textbf{CodeQL Query Generator.}
    This agent composes the generated predicates into a complete query designed to identify vulnerabilities based on the learned patterns. 
    Since CodeQL utilizes a ``select-from-where'' structure similar to SQL, the Query Generator orchestrates the generated predicates and integrates them into a complete query, ensuring they satisfy the structural requirements of a functional query.
    Ultimately, the separation of concerns allows the Predicate Generator to focus on low-level semantic logic, while the Query Generator manages the high-level orchestration and syntactic integration.

\item \textbf{Compilability Checker.}  
    This component utilizes the CodeQL compiler to verify that the generated predicates and queries are syntactically valid and compilable.  
\item \textbf{Syntax Refiner.}  
    If compilation fails, the Syntax Refiner agent is triggered to automatically correct errors. This refinement process is iterative, allowing up to three attempts to achieve successful compilation of the predicates and query.
    
\end{enumerate}

%% file: 4_evaluation.tex
\section{Experimental Setup}
\label{sec:evaluation}

We perform extensive experimental evaluations of the automatic query generation and demonstrate its practical effectiveness in detecting vulnerabilities in real-world Java repositories compared to the standard CodeQL.
Through a series of controlled experiments on a benchmark of known CVEs, we evaluate the performance of LLM-synthesized CodeQL queries. 



\subsection{Data}

In our study, we used CVEs reported in NVD. The main dataset is a collection of CVEs from MoreFixes~\cite{akhoundali2024morefixes}.


\subsubsection{CVE for Pattern Mining}

We utilized the MoreFixes dataset~\cite{akhoundali2024morefixes} as the primary source of VFCs or security patches for our vulnerability pattern mining phase and utilized the CVEs related to the Java programming language in this study.

To ensure the practical relevance of our study, we further refined our selection based on CWE-ID, focusing on the top 15 most dangerous software weaknesses as defined by the MITRE 2025 CWE Top 25~\cite{mitre2025cwetop25}.\footnote{\url{https://cwe.mitre.org/top25/archive/2025/2025_cwe_top25.html}}
Finally, we selected the most recent CVEs available in the MoreFixes database, reserving the latest 20\% from each category as the test set.
We limited our selection to a maximum of 20 CVE-code segment pairs to serve as seeds for the pattern mining process.
However, we found that several of the initial 15 CWE categories were underrepresented in the NVD. Consequently, we narrowed our final scope to 10 CWE IDs that contained a minimum of 20 code segments for pattern extraction. 
The only exception was CWE-434, which provided 19 CVE-code segment pairs; all other selected categories reached the 20-pair threshold.
The details about the data that we used in the vulnerability pattern mining are presented in Table~\ref{tab:cwe_summary}.

\begin{table}[htbp]
\centering
\caption{Distribution of selected CWE categories and experimental data.}
\label{tab:cwe_summary}
\begin{tabular}{p{2cm} 
          p{5cm}       
          p{1.3cm}     
          p{1.3cm} }
\toprule
\textbf{CWE ID} & \textbf{CWE Description} & \textbf{\#CVE Mining Seed} & \textbf{\#CVE Test Data} 
\\ \midrule
CWE-22  & Path Traversal                    & 16 & 9 \\
CWE-78  & OS Command Injection              & 1 & 5 \\
CWE-79  & Cross-site Scripting              & 10 & 37 \\
CWE-89  & SQL Injection                     & 7 & 8 \\
CWE-94  & Code Injection                    & 7 & 8 \\
CWE-352 & Cross-Site Request Forgery        & 10 & 17 \\
CWE-434 & Unrestricted Upload of File with Dangerous Type & 5 & 4 \\
CWE-502 & Deserialization of Untrusted Data & 10 & 15 \\
CWE-787 & Out-of-bounds Write & 6 & 7 \\
CWE-862 & Missing Authorization  & 8 & 2 \\

\midrule
\textbf{Total} & \textbf{10 Categories}     & \textbf{80} & \textbf{112} \\ \bottomrule
\end{tabular}
\end{table}



\subsubsection{CVE Evaluation Dataset}
We extracted a subset of the MoreFixes dataset spanning the same 10 CWE categories used in the pattern mining phase. To construct this test set, we sampled up to 20 of the most recent CVEs per category. 
We applied two primary exclusion criteria to ensure data quality and experimental integrity: (1) each VFC must involve no more than five modified files, and (2) the CVE must not have been included in the initial vulnerability pattern mining phase to prevent data leakage. 



\subsection{Baseline}
For our baseline comparison, we evaluated the detection efficacy of CodeQL when utilizing LLM-generated queries against the default Java security queries provided in the official CodeQL repository.\footnote{https://github.com/github/codeql/blob/main/java/ql/src/Security} 

Additionally, we included PDBERT~\cite{liu2024pre} as a representative baseline based on deep-learning. 
PDBERT is a transformer-based model pre-trained using novel objectives, namely Control Dependency Prediction (CDP) and Data Dependency Prediction (DDP), which can boost the understanding of vulnerable code during fine-tuning.
We selected this approach because it demonstrated superior performance over other contenders in a 2026 comparative study that evaluates vulnerability detection techniques on the repository level~\cite{11408110}.
Crucially, that study included a time-aware evaluation framework to eliminate the risk of data leakage from the test set. 
Notably, PDBERT can be seamlessly adapted to the Java language without requiring the program to be built, mirroring the "build-free" CodeQL configuration used in our experiments.

Regarding a direct LLM-based detection baseline, we omit this approach from our study due to its prohibitive operational costs and limited feasibility for large-scale deployment.
Preliminary experiments showed that scanning a single repository using the budget-friendly Kimi K2.5 model costs 17 USD for a single-shot prompt. 
Scaling this evaluation to over 100 projects benchmark would exceed 1,500 USD, a figures that ignores the necessity of evaluating frontier models, which can be up to six times more expensive, i.e., Sonnet 4.5 cost in OpenRouter is 6 times more expensive than Kimi K2.5.
Furthermore, advanced LLM-based techniques such as VulTrial \cite{widyasari2025let} are similarly cost-inefficient.
While VulTrial reported a cost of 7.46 USD for 435 curated samples, applying that logic to a single repository with 3,719 functions would cost 64 USD using the now-outdated GPT-3.5.
Projecting this across 100 repositories would exceed 6,000 USD for a single scan. Using GPT-4o, the cost per repo jumps to approximately 180 USD, totaling 18,000 USD for the full benchmark. These costs would further escalate if developers of security scanners performed multiple scans or utilized even higher-tier models, making direct LLM detection financially unsustainable compared to the automatic query-generation approach.

Beyond operational costs, omitting direct LLM-based detection inherently protects intellectual property and data confidentiality. 
By avoiding the need to transmit proprietary source code to third-party providers, our methodology sidesteps the legal and security hurdles common in enterprise environments.

While one could argue that deploying open-weights models like Kimi K2.5 locally could mitigate these privacy concerns, the required infrastructure remains inaccessible for most organizations. For context, hosting the full Kimi K2.5 model requires a minimum of four NVIDIA H200 GPUs~\cite{unsloth2025kimi}, with individual units retailing between 30,000 and 40,000 USD. Including the necessary high-performance computing (HPC) setup, the total capital expenditure for local hosting ranges from 150,000 to 300,000 USD~\cite{modal2024h200price}. Such a prohibitive upfront investment is rarely viable for startups or small-to-medium enterprises (SMEs), further validating our approach of using LLMs to generate portable, locally-executable CodeQL queries.



\subsection{LLMs Selection}
In this study, we evaluated a diverse suite of LLMs for the automated synthesis of CodeQL queries. 
Our selection process was primarily guided by the top-performing models in the programming category on OpenRouter \cite{openrouter_programming_2026}. 
We included the three highest-ranked models and supplemented them with representative architectures such as Moonshot Kimi K2.5, which were integrated into the OpenRouter platform on January 27, 2026. 
Furthermore, we incorporated OpenAI GPT 5.2 Codex into our evaluation.
Despite its absence from the current top 10 rankings, OpenAI's widespread adoption ~\cite{sheng2025llms} makes it a critical baseline. 
This selection strategy ensures a broad representation of proprietary models from leading providers, including X, OpenAI, Google, Moonshot AI, Meta, and Anthropic, thereby enhancing the diversity and robustness of our comparative analysis.



\subsection{Implementation Details}
In this experiment, we utilized CodeQL v2.17.3, which supports the ``no-build" feature for project analysis. 
All databases in this study were constructed using the \texttt{--build-mode none} flag. 
This approach ensures that we can evaluate each project without being restricted by specific environment or build-tooling constraints.

For the language model infrastructure, we utilized the OpenRouter API\footnote{\url{https://openrouter.ai/}}, which allowed for precise monitoring of the experiment's monetary costs. 
This setup effectively simulates a real-world production environment where independent hosting of large-scale models is often prohibitive due to the high hardware and maintenance requirements.

\subsection{Evaluation Metrics}

Following previous work~\cite{liiris},
we evaluate the detection performance of the generated queries using three key metrics: Total Detected Vulnerabilities (\#Detected), Average False Discovery Rate (AvgFDR), and the Average F1-score (AvgF1).

We define our evaluation over a dataset $D = \{P_1, \dots, P_n\}$, where each project $P_i$ contains a known set of ground-truth vulnerable program files $V_P^{vul}$. A vulnerability is considered successfully detected if at least one identified file, $File \in Files_P$, intersects with the ground truth:
\begin{equation}
    File \cap V_P^{vul} \neq \emptyset
\end{equation}

The metrics are formally defined as follows. First, we determine the number of valid vulnerable paths for a project $P$:
\begin{equation}
    \#VulFile(P) = |\{File \in Files_P \mid File \cap V_P^{vul} \neq \emptyset\}|
\end{equation}

Project-level Recall is defined as a binary indicator, $Rec(P) = 1$ if $\#VulFile(P) > 0$ and $0$ otherwise. Consequently, the aggregate detection count is given by:
\begin{equation}
    \#Detected(D) = \sum_{P \in D} Rec(P)
\end{equation}

Project-level Precision, $Prec(P)$, is defined as the ratio of valid vulnerable files to the total retrieved files:
\begin{equation}
    Prec(P) = \frac{\#VulFile(P)}{|Files_P|}
\end{equation}

From this, we derive the Average False Discovery Rate across the dataset, which is inversely related to precision:
\begin{equation}
    AvgFDR(D) = \text{avg}_{P \in D, |Files_P|>0} (1 - Prec(P))
\end{equation}

Finally, the Average F1-score (AvgF1) is calculated as the mean harmonic mean of precision and recall across all projects in $D$:
\begin{equation}
    AvgF1(D) = \frac{1}{|D|} \sum_{P \in D} \frac{2 \cdot Prec(P) \cdot Rec(P)}{Prec(P) + Rec(P)}
\end{equation}

We address potential mathematical instabilities by imposing specific constraints on these calculations. Since $Prec(P)$ is undefined when no paths are retrieved ($|Files_P| = 0$), AvgFDR is computed only over the subset of projects where at least one result is produced. In contrast, AvgF1 remains robust across the entire dataset; when no files are detected, $Rec(P) = 0$ naturally forces the F1-score for that project to zero, regardless of the precision value.

%% file: 5_result.tex
\section{Results}
\label{sec:result}

In this section, we present the results to our research questions.

\noindent \textbf{RQ1: LLMs Performance in Generating Compilable \codeql Query} 


\begin{table}[htbp]
\centering
\caption{Model performance in generating compilable predicates and queries without \textit{syntax refiner}.}
\label{tab:result_rq1_query_generation_no_syntax_refiner}
\begin{tabularx}{0.8\columnwidth}{@{}X c@{}}
\toprule
\textbf{Base Model} & \textbf{\# Compilable Queries} \\ \midrule
Gemini 3 Flash Preview         & 17 \\
DeepSeek R1                    & 5 \\
Grok Code Fast 1     & 5 \\
Kimi K2 Thinking     & 0 \\
Kimi K2.5            & 4 \\
Llama 3.3            & 0 \\ 
Minimax M2.1         & 0 \\
Qwen 3 Coder         & 0 \\
GPT 5.2 Codex        & 2\\
Claude Sonnet 4.5           & 0 \\
\midrule
\end{tabularx}
\end{table}

Table~\ref{tab:result_rq1_query_generation_no_syntax_refiner} presents the results of our initial attempt to generate predicates and queries without an additional syntax-refining stage. 
In this experiment, we observed that nearly all models struggled to translate vulnerability patterns into compilable CodeQL code, suggesting that these models lack sufficient familiarity with CodeQL's specific syntactical requirements. 

Consequently, we introduced a \textit{Syntax Refiner Agent} in an attempt to address the syntactic limitations and enhance the overall compilability of the generated queries.
This agent is designed to iteratively correct uncompilable code with a maximum of three repair attempts. 
Furthermore, we explored a hybrid strategy by incorporating Gemini 3 Flash Preview, i.e., the model with the highest success rate in generating compilable queries, as a specialized ``syntax specialist" for more expensive models such as GPT 5.2 Codex and Sonnet 4.5. 
This approach seeks to exploit Gemini 3 Flash Preview's superior syntactical knowledge of CodeQL while simultaneously reducing overall financial costs.


\begin{table}[htbp]
\centering
\caption{Model performance in generating compilable predicates and queries with \textit{syntax refiner}.
}
\label{tab:result_rq1_query_generation_with_syntax_refiner}
\footnotesize
\setlength{\tabcolsep}{4pt}
\begin{tabular}{@{}llrrr@{}}
\toprule
\textbf{Base Model} & \textbf{Syntax Refiner} & \textbf{\#Compilable Queries} & \textbf{Duration (Hour)} & \textbf{Cost (USD)}\\
\midrule
Gemini 3 FP    & = base model & 123 & 8  & 9.25  \\
DeepSeek R1    & = base model & 5   & 18 & 0.00  \\
Grok Code F1   & = base model & 57  & 4  & 3.03  \\
Kimi K2 T      & = base model & 51  & 12 & 12.70 \\
Kimi K2.5      & = base model & 96  & 24 & 12.99 \\
Llama 3.3      & = base model & 0   & 4  & 0.77  \\
Minimax M2.1   & = base model & 11  & 34 & 8.78  \\
Qwen 3 Coder   & = base model & 3   & 3  & 0.76  \\
GPT 5.2 Codex  & Gemini 3 FP  & 105 & 6  & 17.15 \\
Sonnet 4.5     & Gemini 3 FP  & 120 & 14 & 37.10 \\
\bottomrule
\end{tabular}
\end{table}

Based on the experimental results presented in Table~\ref{tab:result_rq1_query_generation_with_syntax_refiner}, we observe that the syntax refiner plays a critical role in the automatic generation of CodeQL queries. 
By providing iterative feedback, it effectively directs the LLM to produce compilable code. 
Furthermore, Gemini 3 Flash Preview's capability for constructing syntactically correct queries was once again demonstrated by a significant increase in success rates. 
Notably, the hybrid approach successfully repaired up to 120 queries that were previously incorrectly composed by Sonnet 4.5, highlighting the value of cross-model syntax refinement.


\begin{rqbox}{Answer to RQ1}
All evaluated LLMs exhibited a lack of syntactic fluency regarding CodeQL's syntax, resulting in low compilation success rates when limited to a single-shot generation. 
To address this bottleneck, an iterative syntax refinement proved highly effective in resolving lexical errors. 
\end{rqbox}

\noindent \textbf{RQ2: Effectiveness of Generated Queries in Detecting Real-World Vulnerabilities} 

Table~\ref{tab:result_rq2_train_all_test_all} summarizes the performance of our automatically generated queries compared to CodeQL's default query set. 
Our experimental results reveal varying strengths among the models. While Claude Sonnet 4.5 achieved the highest vulnerability detection rate at 45.54\%, this performance was offset by a higher FDR. 
In contrast, Kimi K2.5 demonstrated higher balance; it trailed Sonnet 4.5 by only a single CVE in terms of detection but maintained a FDR that was more than three percentage points lower. Consequently, Kimi K2.5 achieved the highest overall average F1-score of 24.04\%. This represents an 82\% improvement of average F1-Score over the CodeQL baseline, suggesting that our automated synthesis approach is a more practical alternative for real-world deployment. 

Interestingly, Gemini 3 Flash Preview, which previously exhibited superior proficiency in generating compilable queries, failed to correctly translate the underlying logic of vulnerability patterns into semantically meaningful CodeQL queries. 
This performance gap suggests that while Gemini 3 is highly effective as a \textit{syntax refiner}, it lacks the deeper reasoning capabilities required to act as a primary \textit{logic reasoner} for complex query synthesis.

\begin{rqbox}{Answer to RQ2}
While Claude Sonnet 4.5 leads in raw discovery, Kimi K2.5 achieves a superior precision-recall balance, outperforming the standard CodeQL suite by 82\% in F1-score. This demonstrates that LLM-synthesized queries effectively surpass the baselines in real-world applications.
\end{rqbox}

\begin{table*}[t]
\centering
\caption{Model performance in detecting real-world vulnerabilities.}
\label{tab:result_rq2_train_all_test_all}
\begin{tabularx}{\textwidth}{@{}
  >{\raggedright\arraybackslash}X
  >{\centering\arraybackslash}X
  >{\centering\arraybackslash}X
  >{\centering\arraybackslash}X
  >{\centering\arraybackslash}X@{}}
\toprule
\textbf{Model} & \textbf{\#Detected (/112) ($\uparrow$)} &  \textbf{Detection Rate (\%) ($\uparrow$)} & \textbf{Avg FDR (\%) ($\downarrow$)} & \textbf{Avg F1 Score} ($\uparrow$) \\ \bottomrule
CodeQL   & 19 & 16.96 & 89.19 & 13.20 \\ 
PDBERT   & 6  & 5.36  & 43.06 &  12.06 \\ \bottomrule
Gemini 3 FP         & 3 & 2.68 & 79.48 & 4.74 \\
Grok Code Fast 1    & 25  & 22.32 & 95.20 & 7.90 \\
Kimi K2.5           & 50 & 44.64 & 89.92 & \textbf{24.04} \\
Minimax M2.1        & 8 & 7.14 & 93.43 & 6.84 \\
GPT 5.2 Codex + Gemini 3 FP  & 27 & 24.11  & 83.84 & 21.47  \\
Claude Sonnet 4.5 + Gemini 3 FP  &  51 & 45.54  & 93.13 & 11.94 \\
\bottomrule
\end{tabularx}
\end{table*}

\noindent \textbf{RQ3: Where Automated Generation Performs Better} 

In an effort to further optimize our automated generation process, we conducted additional experiments to explore the impact of patch complexity on model performance. 
Based on the intuition that LLMs may distill information more effectively from concise contexts, we investigated whether restricting the input to CVEs with ``short fixes", defined as changes involving no more than 10 lines within a single file, improves the quality of the generated queries. For this phase, we selected the three highest-performing models from our initial evaluations: Kimi K2.5, GPT 5.2 Codex (paired with Gemini 3 Flash), and Sonnet 4.5 (paired with Gemini 3 Flash Preview). 

Table~\ref{tab:result_rq3_train_short_test_all} summarizes the results of mining vulnerability patterns from short CVEs when evaluated against the full test suite. 
This setup tests whether models can better identify the core characteristics of a vulnerability when provided with highly focused context. 
Our results indicate that Sonnet 4.5 was the only model to benefit from this restriction, increasing its Avg F1-score by 63\% (from 11.94\% to 19.49\%). Conversely, Codex and Kimi K2.5 experienced significant performance degradation, suggesting that these models leverage the broader context of larger patches to craft more effective queries.

Subsequently, we evaluated the performance of the generated queries specifically on CVEs with short changes, as presented in Tables~\ref{tab:result_rq3_train_short_test_short} and \ref{tab:result_rq3_train_all_test_short}. 
This configuration explores whether detecting atomic vulnerabilities is inherently easier. 
This inquiry is particularly rigorous because the limited ground truth in short changes severely penalizes the detection rate; a meaningful query must hit one exact file to be counted as successful.
Our findings confirm that the generated queries are indeed semantically precise, as evidenced by the low AvgFDR and high detection rate achieved by Codex. This phenomenon demonstrates that Codex successfully distilled the ``essence" of the vulnerabilities into precise logic, achieving a detection rate significantly higher than the CodeQL baseline. Notably, the highest overall Avg F1-score was achieved by utilizing GPT 5.2 Codex to mine patterns from the full dataset and then applying them to detect vulnerabilities in short-change CVEs, i.e., intra-procedural vulnerabilities.
\ly{As for the term of "intra-procedural vulnerabilities", it is better to explain or setup in the RQ3 discussion. Since as for some readers, they may not know what is "intra-procedural vulnerabilities"} \iv{updated in the RQ3 explanation}

\begin{rqbox}{Answer to RQ3}
Using all available information from a vulnerability report is the most effective way to generate a query. However, these queries are most helpful for catching small, localized bugs, which are characteristic of intra-procedural vulnerabilities.
\end{rqbox}

\begin{figure}[hb!]
    \centering
    \includegraphics[width=\columnwidth]{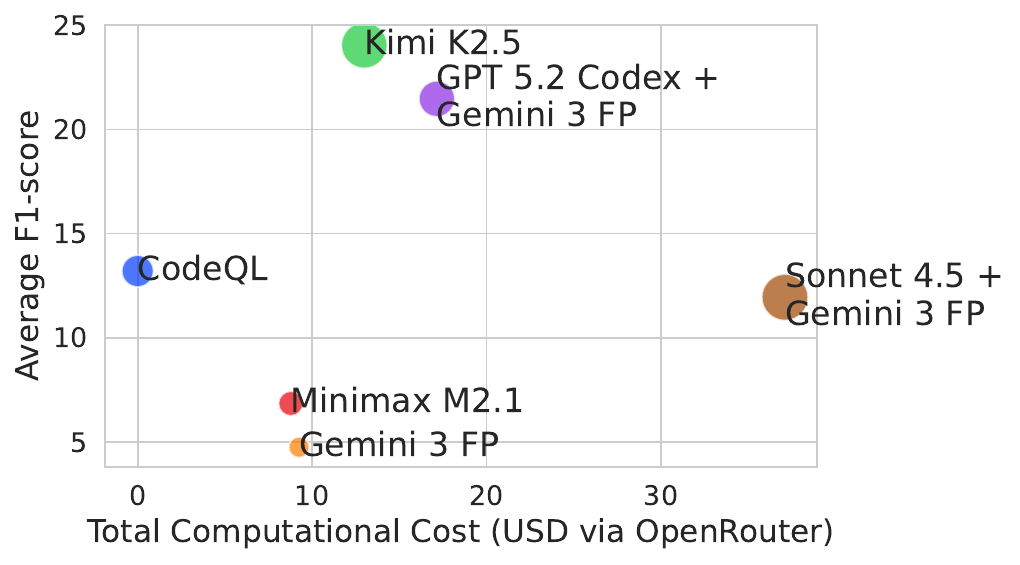}
    \caption{Cost-efficiency visualization for automated CodeQL query generation. \textbf{The area of each marker is proportional to the \#Detected}.}
    \label{fig:cost_performance}
\end{figure}

\begin{table*}[ht!]
\centering
\caption{Model performance in detection of real-world vulnerabilities - vuln pattern seed: short CVEs, test: all.}
\label{tab:result_rq3_train_short_test_all}
\begin{tabularx}{\textwidth}{@{}
  >{\raggedright\arraybackslash}X
  >{\centering\arraybackslash}X
  >{\centering\arraybackslash}X
  >{\centering\arraybackslash}X
  >{\centering\arraybackslash}X@{}}
\toprule
\textbf{Model} & \textbf{\#Detected (/112) ($\uparrow$)} &  \textbf{Detection Rate (\%) ($\uparrow$)} & \textbf{Avg FDR (\%) ($\downarrow$)} & \textbf{Avg F1 Score} ($\uparrow$) \\ \midrule
CodeQL (Baseline)   & 19 & 16.96 & 89.19 & 13.20 \\ \midrule
Kimi K2.5           & 17 & 15.18 & 86.12 & 14.50 \\
GPT 5.2 Codex + Gemini 3 FP  & 2 & 1.79  & 94.57 & 2.69  \\
Claude Sonnet 4.5 + Gemini 3 FP  &  \textbf{31} & \textbf{27.68} & \textbf{84.96} & \textbf{19.49}\\
\midrule
\end{tabularx}
\end{table*}

\begin{table*}[ht!]
\centering
\caption{Model performance in detection of real-world vulnerabilities - vuln pattern seed: short CVEs, test: short CVEs.}
\label{tab:result_rq3_train_short_test_short}
\begin{tabularx}{\textwidth}{@{}
  >{\raggedright\arraybackslash}X
  >{\centering\arraybackslash}X
  >{\centering\arraybackslash}X
  >{\centering\arraybackslash}X
  >{\centering\arraybackslash}X@{}}
\toprule
\textbf{Model} & \textbf{\#Detected (/25) ($\uparrow$)} &  \textbf{Detection Rate (\%) ($\uparrow$)} & \textbf{Avg FDR (\%) ($\downarrow$)} & \textbf{Avg F1 Score} ($\uparrow$) \\ \midrule
CodeQL (Baseline)   & 5 & 20 & 92.95 & 10.43 \\ \midrule
Kimi K2.5           & 7 & 28 & 87.44 & 23.53 \\
GPT 5.2 Codex + Gemini 3 FP  & 1 & 4  & \textbf{75} & 6.90  \\
Claude Sonnet 4.5 + Gemini 3 FP  &  \textbf{10} & \textbf{32}  & 80.39 & \textbf{24.32} \\
\midrule
\end{tabularx}
\end{table*}

\begin{table*}[htbp!]
\centering
\caption{Model performance in detection of real-world vulnerabilities - vuln pattern seed: all, test: short CVEs.}
\label{tab:result_rq3_train_all_test_short}
\begin{tabularx}{\textwidth}{@{}
  >{\raggedright\arraybackslash}X
  >{\centering\arraybackslash}X
  >{\centering\arraybackslash}X
  >{\centering\arraybackslash}X
  >{\centering\arraybackslash}X@{}}
\toprule
\textbf{Model} & \textbf{\#Detected (/25) ($\uparrow$)} &  \textbf{Detection Rate (\%) ($\uparrow$)} & \textbf{Avg FDR (\%) ($\downarrow$)} & \textbf{Avg F1 Score} ($\uparrow$) \\ \midrule
CodeQL (Baseline)   & 5 & 20 & 92.95 & 10.43 \\ 
PDBERT   & 3 & 12 & 50 & 19.35 \\
\midrule
Kimi K2.5           & 10 & 40 & 89.68 & 16.41 \\
GPT 5.2 Codex + Gemini 3 FP  & 10 & 40  & \textbf{75.31} & \textbf{30.53} \\
Claude Sonnet 4.5 + Gemini 3 FP  &  \textbf{12} & \textbf{48}  & 94.50 & 9.87 \\
\midrule
\end{tabularx}
\end{table*}

\noindent \textbf{RQ 4: Cost-Effectiveness and Practicality of Automatic Query Generation}

Figure~\ref{fig:cost_performance} illustrates the trade-off between computational cost and detection performance across the evaluated models. 
While Kimi K2.5 does not achieve the absolute highest number of detections (\#Detected), it yields the superior Average F1-score at a highly efficient cost point. 
Specifically, the model generated 96 CodeQL queries for less than 13 USD, successfully detecting 50 CVEs within the benchmark. This represents a 263\% increase in the detection rate over the CodeQL baseline, rising from 19 to 50 detected vulnerabilities.
These results suggest that Kimi K2.5 serves as a highly cost-effective alternative for automatically synthesizing CodeQL queries from vulnerability reports to bolster software security.

Our experiments involving the detection of CVEs with short fixes further underscore the practical utility of Kimi K2.5. From both a budgetary and performance perspective, it represents the most viable LLM for real-world vulnerability pattern mining; it consistently outperforms the standard CodeQL baseline while remaining 65\% more affordable than Sonnet 4.5 and 25\% cheaper than GPT 5.2 Codex. This dual efficiency in cost and performance confirms its suitability for practical deployment, particularly for automated learning tasks aimed at maintaining and updating CodeQL query packages in rapidly evolving security environments.

\begin{rqbox}{Answer to RQ4}
Kimi K2.5 represents the most viable LLM for practical deployment, offering an optimal balance between computational budget and detection performance. It achieved a superior Average F1-score by synthesizing 96 CodeQL queries for under USD 13, successfully detecting 50 CVEs. Furthermore, Kimi K2.5 is 65\% more affordable than Claude Sonnet 4.5 and 25\% cheaper than GPT 5.2 Codex, while consistently outperforming the CodeQL baseline.
\end{rqbox}

%% file: 6_lesson_learned.tex
\section{Discussion}\label{sec:discussion}
\subsection{False Positive Observation}
While LLM-generated queries outperformed CodeQL in detection rates while maintaining comparable False Discovery Rates (FDR), we observed a notable volume of false alarms. To investigate potential remediations, we analyzed the query derived from the fix for CVE-2022-40152, which achieved high recall but low precision. This query, presented in Listing~\ref{lst:cve_query}, was designed to identify recursive method calls lacking both depth parameters and guard conditions. The LLM successfully captured the essential logic for identifying recursion and defined two predicates to act as recursion-depth checkers; however, due to space limitations, some auxiliary details are omitted from the listing.

\begin{figure}[h]
\begin{lstlisting}[caption={LLM-generated CodeQL query for CVE-2022-40152.}, label={lst:cve_query}]
import java
class RecursiveMethod extends Method {
  RecursiveMethod() {...}
  }
  MethodCall getARecursiveCall() {...}
}
predicate hasDepthTrackingParameter(RecursiveMethod m) {...}
predicate hasDepthLimitGuard(RecursiveMethod m) {...}
from RecursiveMethod m
where not hasDepthTrackingParameter(m) and not hasDepthLimitGuard(m)
select m, "Recursive method lacks depth limiting mechanisms, potentially leading to stack exhaustion (CWE-787)."
\end{lstlisting}
\label{listing:false-positive}
\end{figure}

Despite the seemingly sound logic, this query flagged over 500 files, prompting a manual review of the alerts. Analysis of a representative false positive in the novel-plus repository~\footnote{\url{https://github.com/201206030/novel-plus/blob/f3f37721b119820f95e4dd9d8643e03355085d32/novel-admin/src/main/java/com/java2nb/common/utils/TimeUtils.java\#L158}} revealed that the flagged segment was unreachable ``dead code''. 

Although the CodeQL alert was technically accurate, as logic poses a legitimate risk if executed, the specific code segment is never called within the application.
Theoretically, the method triggers an out-of-bounds error when called with \texttt{Integer.MAX\_VALUE} as `day` parameter. 
However, because this path is inactive, it poses no real-world exploit risk and is therefore categorized as a false positive. This finding suggests that future research should integrate query generation with reachability analysis to filter out inactive code paths and reduce manual verification efforts. Such filtering is particularly valuable given that dead code is a widespread phenomenon in Java applications~\cite{caivano2021exploratory, caivano2023spread}, and its presence systematically inflates false positive rates in static analysis pipelines.




\subsection{Lessons Learned and Implications}\label{sec:implications}
\begin{figure*}[htb]
    \centering
    \includegraphics[width=\textwidth]{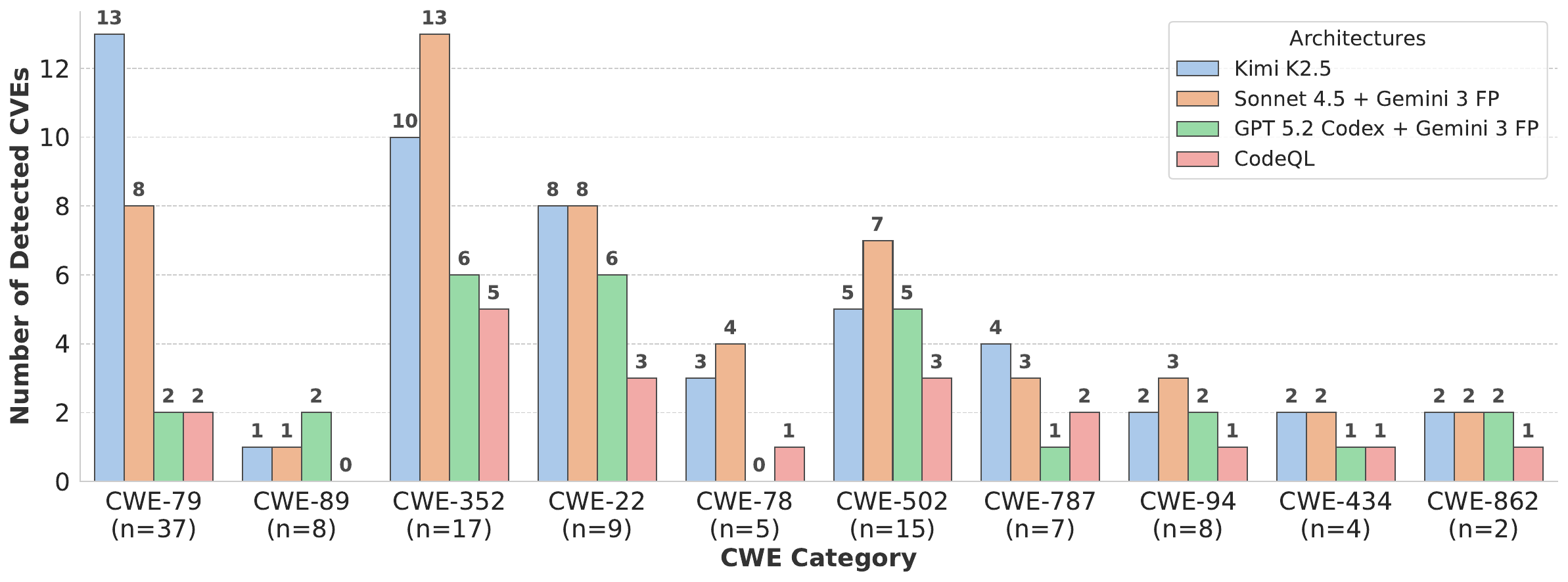}
    \caption{Comparative analysis of detection counts for the top 10 CWE classes. 
    }
    \label{fig:multi-model-cwe}
\end{figure*}

\subsubsection*{\textbf{Developers cannot rely solely on the prepackaged queries in CodeQL}} 

Despite its widespread adoption and endorsement by industry leaders such as Microsoft~\cite{microsoft2023powerpages} and GitHub~\cite{github_codeql_about}, CodeQL's efficacy in detecting modern, high-impact vulnerabilities remains notably constrained. 
Our experiments, which were conducted on a diverse subset of vulnerabilities from the last five years, reveal a detection rate of less than 20\% when using standard CodeQL query suites. 
This performance suggests that while the framework provides a robust engine for semantic analysis, the official, prebuilt query suites struggle to maintain pace with an increasingly sophisticated and rapidly evolving software vulnerabilities. 
Consequently, achieving comprehensive detection coverage remains heavily dependent on the manual, time-intensive curation of custom queries by security experts.

\noindent \subsubsection*{\textbf{While evaluated LLMs consistently outperform the standard CodeQL baseline, our results indicate that no single model dominates all vulnerability categories}}
Figure~\ref{fig:multi-model-cwe} presents a comparative analysis of detection capabilities categorized by CWE ID. Across all categories, both the Sonnet 4.5 and Gemini 3 Flash Preview hybrid, as well as Kimi K2.5, consistently outperform standard CodeQL queries in vulnerability detection. These results underscore the significant potential of leveraging LLMs to automatically mine vulnerability patterns from reported CVEs, thereby bolstering repository resilience against dangerous software vulnerabilities.

Furthermore, our findings suggest that LLM configurations can be optimized based on specific security requirements and vulnerability classes. 
For instance, 
XSS vulnerabilities (CWE-79) are most effectively addressed using Sonnet 4.5 as the base model for pattern mining. Conversely, Kimi K2.5 demonstrates superior efficacy in handling Cross-Site Request Forgery (CSRF), OS Command Injection, Deserialization of Untrusted Data, and Code Injection.

These performance variations likely stem from the distinct data distributions utilized during each model’s pre-training phase. 
Although the specific composition of proprietary training datasets is rarely disclosed by providers, our study provides empirical evidence identifying which models excel at detecting specific vulnerability classes. 
These findings offer valuable insights into the relative strengths of each model across diverse security categories. 
For example, projects that utilize SQL integration are susceptible to CWE-89 (SQL Injection) vulnerabilities; therefore, they might benefit more from mining vulnerability patterns using Kimi K2.5.

Notably, all LLM-generated queries successfully detected every vulnerability in the CWE-862 test set, despite the limited amount of historical training data available for the models. 
In practice, maintaining a multi-model ensemble can maximize detection rates across diverse CWE IDs, ultimately strengthening the overall security posture of the software.

\subsubsection*{\textbf{Function-level analysis may not fully capture repository-level vulnerabilities}}
In our experiments, PDBERT performed worse than CodeQL. This happened even though we made the evaluation easier for PDBERT by only analyzing specific functions and using a fairly balanced sample from the MoreFixes dataset (1010 vulnerable functions and 1223 non-vulnerable functions).
Furthermore, we also computed the precision, recall, and Avg F1-Score on the file level, identical to the CodeQL evaluation.

One of the reasons for this is that not all vulnerabilities can be simplified down to a single function. 
For example, CVE-2022-46688, a CSRF vulnerability in Jenkins—was fixed by adding a @RequirePOST annotation rather than changing the code logic inside the function. 
When a tool only looks at the function itself, it likely ignores this annotation and fails to recognize it as the root cause of the vulnerability.

\subsubsection*{\textbf{Real challenge is in filtering false positive}}
Our experiments demonstrate that LLM-generated queries significantly improve vulnerability detection rates compared to standard CodeQL queries. 
As our study demonstrates a significant leap in recall, future research can build upon this foundation by prioritizing advanced false-positive filtering to refine the precision of LLM-generated queries. 
Such advancements would likely lead to a higher average F1-score, effectively bridging the gap between high detection capabilities and the need for reduced manual verification in automated security analysis.

%% file: 7_related_works.tex
\section{Related Work}\label{sec:related}
A highly-significant research area that converts textual information into a query is the text-to-SQL study~\cite{hong2025next}. 
The evolution of text-to-SQL research has progressed from early rule-based and template-driven systems ~\cite{li2014constructing, yu-etal-2018-spider}, which lacked the scalability to handle linguistic diversity, to sophisticated deep neural architectures. 
While initial sequence-to-sequence models like Seq2SQL~\cite{zhong2017seq2sql} and RYANSQL~\cite{choi2021ryansql} introduced sketch-based generation to improve generalization, they frequently struggled with complex constructs like nested subqueries. 
This led to the adoption of pre-trained language models (PLMs) such as BERT~\cite{devlin2019bert} and RoBERTa~\cite{liu2019roberta}, which, when fine-tuned, significantly enhanced structural comprehension~\cite{zhong2017seq2sql}. 
Despite these advancements, the high cost of task-specific fine-tuning and limited cross-domain adaptability remained persistent challenges. The current era is instead defined by LLMs such as GPT-4 and LLaMA~\cite{touvron2023llama}, shifting research toward prompt engineering, in-context learning, and chain-of-thought reasoning. 
Modern frameworks like SQL-PaLM~\cite{sun2023sql} and StructGPT~\cite{jiang2023structgpt} leverage these capabilities to achieve state-of-the-art performance in complex SQL generation without exhaustive retraining.

Building on this momentum, recent security research has introduced IRIS~\cite{liiris}, a neuro-symbolic framework that uses LLMs to infer CWE-specific source and sink labels for third-party library APIs, which are then integrated into a CodeQL engine to identify vulnerability paths.
While IRIS enhances static analysis by automating specification discovery, it still relies on predefined query templates that only support taint analysis vulnerability detection. 
In contrast, we examine whether LLMs can move beyond simple labeling to autonomously generate complete, executable CodeQL queries by bypassing rigid templates and domain-specific fine-tuning.

Additionally, recent research has highlighted the limitations of function-centric detection by shifting the focus toward repository-level contexts \cite{11408110}.
VulEval's authors demonstrate that capturing cross-functional dependencies significantly improves detection precision and recall compared to traditional intra-procedural methods. 
While VulEval emphasizes the importance of repository-wide context for evaluating deep learning models, it primarily focuses on identifying existing vulnerabilities within a graph-based or sequence-based representation and has been experimented on C/C++ projects. In contrast, our work leverages Java repository-level information not just for detection, but as a semantic foundation for autonomously synthesizing CodeQL queries, enabling the transformation of high-level repository context into executable security logic.

%% file: 8_future_works.tex
\section{Conclusion and Future Work}\label{sec:conclusion}

In this work, we provide empirical evidence that leveraging LLMs to recognize and translate vulnerability patterns from NVD entries significantly improves detection rates without compromising precision. 
Our experiments demonstrate that patterns mined via Kimi K2.5 and translated into CodeQL queries achieve a 263\% improvement in detection performance compared to baseline methods. Notably, this methodology mitigates data confidentiality concerns, as it eliminates the need to submit project-specific source code to third-party LLM providers. 

While repeated LLM inference can be cost-prohibitive, we found that Kimi K2.5 offers an optimal balance between CodeQL syntax proficiency and reasoning capability. Furthermore, for tasks requiring deeper logic, such as the initial vulnerability pattern mining, we observed that high-reasoning models such as GPT 5.2 Codex or Claude Sonnet 4.5 can be effectively paired with Gemini 3 Flash Preview. 
In this hybrid configuration, the latter serves as a cost-effective syntax corrector, significantly reducing operational expenses compared to utilizing standalone frontier models for both initial generation and refinement.

Looking ahead, we intend to investigate the cross-language transferability of LLM-mined patterns. 
For instance, adapting C/C++ vulnerability logic to detect similar flaws in Java projects. 
Such research is critical because existing CVE datasets are predominantly composed of C/C++ entries, while widely used languages like Python and Java have a smaller fraction of reported vulnerabilities. 
This data imbalance presents a valuable opportunity to explore cross-language vulnerability transfer, potentially leading to more robust CodeQL representations that are less dependent on language-specific datasets.

%% file: 9_data_availability.tex
\section*{Data Availability}\label{sec:data}

For transparency and reproducibility, we have made our study's replication package, including all associated code and scripts, publicly available at \url{https://figshare.com/s/b4cc715fa3415a52bad5?file=63158995} (DOI: 10.6084/m9.figshare.31866559).
Additionally, the raw datasets used for query generation and performance evaluation are archived at \url{https://figshare.com/s/523e6adee7278c8dadf6} (DOI: 10.6084/m9.figshare.31866628).
